\documentclass[twocolumn,english,aps,prb,twocolum,superscriptaddress,bibnotes,amsmath,amssymb,floatfix]{revtex4-2}
\usepackage[colorlinks=true,citecolor=blue,linkcolor=magenta]{hyperref}

\usepackage[markup=nocolor, authormarkupposition=left]{changes}
\usepackage[english]{babel}
\usepackage{amsmath,amsfonts,amssymb}
\usepackage{url}

\usepackage{siunitx}
\usepackage{graphicx}
\graphicspath{{./Figures/}}

\usepackage{changes}
\usepackage{soul}
\begin{document}

%\title{Photonic integrated lithium niobate based lasers for coherent ranging}
%\title{Ultrafast tunable photonic integrated lithium niobate based lasers}
%\title{Ultrafast tunable Lithium niobate integrated photonics based lasers}
%\title{Ultrafast tunable lasers based on lithium niobate integrated photonics}
\title{Ultrafast tunable lasers using lithium niobate integrated photonics}
	
\author{Viacheslav Snigirev}
\thanks{These authors contributed equally to this work.}
\affiliation{Institute of Physics, Swiss Federal Institute of Technology Lausanne (EPFL), CH-1015 Lausanne, Switzerland}
	
\author{Annina Riedhauser}
\thanks{These authors contributed equally to this work.}
\affiliation{IBM Research Europe, Zurich, Säumerstrasse 4, CH-8803 Rüschlikon, Switzerland}

\author{Grigory Lihachev}
\thanks{These authors contributed equally to this work.}
\affiliation{Institute of Physics, Swiss Federal Institute of Technology Lausanne (EPFL), CH-1015 Lausanne, Switzerland}
	
\author{Johann Riemensberger}
%\thanks{These authors contributed equally to this work.}
\affiliation{Institute of Physics, Swiss Federal Institute of Technology Lausanne (EPFL), CH-1015 Lausanne, Switzerland}
	
\author{Rui~Ning~Wang}
\affiliation{Institute of Physics, Swiss Federal Institute of Technology Lausanne (EPFL), CH-1015 Lausanne, Switzerland}

\author{Charles M\"ohl}
\affiliation{IBM Research Europe, Zurich, Säumerstrasse 4, CH-8803 Rüschlikon, Switzerland}
	
\author{Mikhail Churaev}
\affiliation{Institute of Physics, Swiss Federal Institute of Technology Lausanne (EPFL), CH-1015 Lausanne, Switzerland}

\author{Anat Siddharth}
\affiliation{Institute of Physics, Swiss Federal Institute of Technology Lausanne (EPFL), CH-1015 Lausanne, Switzerland}

\author{Guanhao Huang}
\affiliation{Institute of Physics, Swiss Federal Institute of Technology Lausanne (EPFL), CH-1015 Lausanne, Switzerland}
	
\author{Youri Popoff}
\affiliation{IBM Research Europe, Zurich, Säumerstrasse 4, CH-8803 Rüschlikon, Switzerland}
\affiliation{Integrated Systems Laboratory, Swiss Federal Institute of Technology Zurich (ETH Z\"{u}rich), CH-8092 Z\"{u}rich, Switzerland}

\author{Ute~Drechsler}
\affiliation{IBM Research Europe, Zurich, Säumerstrasse 4, CH-8803 Rüschlikon, Switzerland}

\author{Daniele Caimi}
\affiliation{IBM Research Europe, Zurich, Säumerstrasse 4, CH-8803 Rüschlikon, Switzerland}
	
\author{Simon H\"onl}
\affiliation{IBM Research Europe, Zurich, Säumerstrasse 4, CH-8803 Rüschlikon, Switzerland}
	
\author{Junqiu Liu}
\affiliation{Institute of Physics, Swiss Federal Institute of Technology Lausanne (EPFL), CH-1015 Lausanne, Switzerland}	
	
\author{Paul Seidler}
\email{pfs@zurich.ibm.com}
\affiliation{IBM Research Europe, Zurich, Säumerstrasse 4, CH-8803 Rüschlikon, Switzerland}
	
\author{Tobias J. Kippenberg}
\email{tobias.kippenberg@epfl.ch}
\affiliation{Institute of Physics, Swiss Federal Institute of Technology Lausanne (EPFL), CH-1015 Lausanne, Switzerland}

	%\begin{abstract}
	%The lithium niobate (LN) photonics experienced a considerable breakthrough with wafer-scale fabrication of thin-film lithium niobate on-insulator (LNOI). The lithium niobate integration opens up new perspectives in electro-optic modulation, frequency shifting, second harmonic generation and optical routing. The low-loss integration of $\chi^{(2)}$ materials is crucial for the quantum applications as well, including single photon transduction for quantum interconnects. However, the conventional integrated lithium niobate devices suffer from sufficient optical losses, including insertion losses to the chip and linear loss of the wavegudies. Here we demonstrate integration of thin-film LNOI with an ultra-low loss silicon nitride platform.
	%\end{abstract}
\maketitle

\textbf{%The Pockels effect associated with ferro-electric materials such as Lithium Niobate
	%$\mathrm{LiNbO_3}$ is the basis of electro-optic high speed modulators that are ubiquitous in modern optical telecommunications.
	Recent advances in the processing of thin-film lithium niobate (LiNbO$\boldsymbol{_3}$) on insulator (LNOI) have enabled low-loss photonic integrated circuits \cite{Zhang:17, krasnokutska2018}, modulators with improved half-wave voltage \cite{Wang2018,He2019}, electro-optic frequency combs \cite{Zhang2019} and novel on-chip electro-optic devices, with applications ranging from 5G telecommunication \cite{yang20194} and microwave photonics \cite{marpaung2019integrated} to microwave-to-optical quantum interfaces \cite{Lambert2020, Javerzac-Galy2016}.
	Lithium niobate integrated photonic circuits could equally be the basis of integrated frequency-agile lasers with narrow linewidth.
	Pioneering work on polished lithium niobate crystal resonators \cite{PhysRevLett.92.043903,1372608,Soltani:16} has led to the development of electrically tunable narrow-linewidth lasers \cite{dale2014ultra}. %, suitable for coherent laser ranging. 
	% Yet, these devices have not been shown to date using lithium niobate integrated photonic circuits.
	Here we report low-noise frequency-agile lasers based on lithium niobate integrated photonics and demonstrate their use for coherent laser ranging. This is achieved through heterogeneous integration of ultra-low-loss silicon nitride (Si$\boldsymbol{_3}$N$\boldsymbol{_4}$) photonic circuits \cite{liu2021high} with thin-film lithium niobate via direct wafer bonding. The hybrid platform  features low propagation loss of 8.5 dB/m enabling narrow-linewidth lasing (intrinsic linewidth of 3~kHz) by self-injection locking to a III--V semiconductor laser diode. The hybrid mode of the resonator allows electro-optical laser frequency tuning at a speed of 12~petahertz/s ($12\times 10^{15}$~Hz/s) with high linearity, low hysteresis and while retaining narrow linewidth.
	Using this hybrid integrated laser, we perform a proof-of-concept frequency-modulated continuous-wave (FMCW) LiDAR ranging experiment, with a resolution of 15 cm.  By fully leveraging the high electro-optic coefficient of lithium niobate, with further improvements in photonic integrated circuits design, these devices can operate with CMOS-compatible voltages, or achieve millimetre-scale distance resolution. Endowing low loss silicon nitride integrated photonics with lithium niobate, gives a platform with wide transparency window, that can be used to realize  
	%Given the wide transparency window, the hybrid platform can be used to realize 
	ultrafast tunable lasers from the visible to the mid-infrared, with applications ranging from optical coherence tomography (OCT) and LiDAR to environmental sensing.}

%Lithium niobate is an attractive material for electro-optic devices, and has been widely used for many decades \cite{poberaj2012lithium}. 
Lithium niobate is an attractive material for electro-optic devices, and has been widely used for many decades. 
It exhibits a transparency window from ultraviolet to mid-infrared wavelengths and has a large Pockels coefficient \cite{Jhans1986,Weis1985}, which is key for efficient, low-voltage, and high-speed modulation. 
Integrated photonics based on materials exhibiting the Pockels effect --- such as AlN \cite{Xiong2012} and GaP \cite{wilson2020integrated} --- have been demonstrated before, but only recently for lithium niobate \cite{Wang:14,Wang:15}. 
%Over the past years, following early attempts to create integrated photonics based on lithium niobate \cite{Wang:14,Wang:15}, a variety of devices have been demonstrated \cite{Zhang:17,krasnokutska2018}. 
Following the commercial availability of lithium niobate on insulator via wafer bonding and smart-cut, there has also been substantial progress in the etching of low-loss lithium niobate waveguides, culminating in linear optical loss of  2.7 dB/m and ring resonators with intrinsic Q-factor of $10\times 10^6$~\cite{Zhang:17}. 
The majority of works have utilized argon ion beam etching to manufacture ridge waveguide structures, which enabled modulators operating at CMOS voltages \cite{Wang2018}, high-efficiency QPSK modulators \cite{xu2020high}, as well as electro-optic frequency combs \cite{Zhang2019}. 
%The majority of work has utilized argon ion beam etching to manufacture ridge waveguide structures \cite{siew2018ultra}, which enabled modulators operating at CMOS voltages \cite{Wang2018}, high efficiency QPSK modulators \cite{xu2020high}, as well as electro-optic frequency combs \cite{Zhang2019}. 
In addition, the platform has provided a route to creating interfaces using cavity electro-optics that efficiently couple microwave to optical fields \cite{Javerzac-Galy2016,Lambert2020}. 
Further potential applications include the creation of synthetic dimensions for topological photonics \cite{hu2020realization}.
In addition to direct etching, recently also heterogeneous integration \cite{komljenovic2018photonic, plossl1999wafer} of lithium niobate chiplets onto silicon nitride \cite{chang2017heterogeneous} or silicon (Si) \cite{Weigel2016} integrated photonic circuits has been demonstrated.

%One advantage of this heterogeneous integration approach is that one can utilize mature integrated low loss photonic circuits for which numerous integrated photonic building blocks exist, and evade precise and low roughness etching of $\mathring{LiNbO_3}$. 

Beyond application for electro-optical modulators, an integrated photonics platform with a large Pockels coefficient and low propagation loss fulfils all the requirements for realizing integrated frequency-agile laser sources, which feature fast, linear and mode-hop-free tuning. 
Specifically, such frequency-agile electro-optical lasers can be realized by self-injection locking of a III--V diode laser to an high-Q optical cavity \cite{kondratiev2017self}. 
This technique, in conjunction with low-loss integrated photonic resonators, has already enabled compact chip-scale lasers with Lorentzian linewidth on the order of one hertz using silicon nitride \cite{Jin2021}. 
Yet to date, no integrated hybrid lasers based on materials displaying Pockels effect have been demonstrated.
In contrast to thermal or piezoelectrical \cite{tian2020hybrid,hosseini2015stress} tuning of optical microresonators, electro-optical materials can enable ultra-fast optical tuning with theoretically tens of GHz intracavity modulation bandwidth. 
%while retaining narrow optical linewidth operation.
Integrated electro-optical materials can therefore be used to realize a host of laser structures, such as widely tunable Vernier lasers \cite{fan2016optically} or fast, mode-hop-free and continuously tunable lasers for a multitude of applications that include coherent LiDAR \cite{bostick1967carbon,martin2018photonic,rogers2021universal}, trace-gas spectroscopy \cite{millot2016frequency,debecker2005high}, and optical coherence tomography (OCT) \cite{ji2019chip}. 	   

Here, we demonstrate for the first time a self-injection-locked narrow-linewidth frequency-agile laser based on a lithium niobate integrated photonic circuit. 
This is achieved on a heterogeneous integrated lithium niobate platform combining low-loss silicon nitride photonic waveguides with lithium niobate by wafer-scale bonding \cite{churaev2021}. Damascene silicon nitride waveguides feature tight confinement, ultra-low propagation loss ($<$1~dB/m), and high-power handling \cite{liu2021high}. They can be manufactured with high yield on a wafer scale and are already available from a commercial foundry.
%This is achieved using a hybrid platform approach \cite{churaev2021}, that heterogeneously combines thin-film lithium niobate with tight confinement, ultra-low loss ($<$1~dB/m), and high-power handling of Damascene $\mathrm{Si_3N_4}$ photonic circuits \cite{Brasch:15}, that can be manufactured with high yield on a wafer scale \cite{liu2021high} and are already available via a commercial foundry. 
Additional advantages of the silicon nitride waveguide platform include low gain of Raman and Brillouin nonlinearities \cite{Karpov:16, Gyger:20} and radiation hardness \cite{Brasch:14}.
Our heterogeneous integrated lithium niobate on Damascene silicon nitride (LNOD) platform enables high-Q resonances (loaded linewidth 100~MHz), provides a high yield of bonded devices, and exhibits low (compared to ridge LNOI) insertion loss (3.9~dB per facet \cite{churaev2021}). 
Combining the unique properties of both platforms enables self-injection locking with laser frequency noise reduction of 20~dB, a fundamental linewidth of 3~kHz, and frequency-agility with a flat actuation bandwidth up to 100~MHz. Using this hybrid integrated laser source, we perform FMCW LiDAR experiments in lab environment with resolution of 15 cm. %, with the hybrid integrated laser output power as high as 0.1 mW and. 

% In contrast to recently demonstrated hybrid integrated lasers using $\mathrm{AlN}$ actuators \cite{lihachev2021ultralow}, the  LiNbO$_3$-Si$_3$N$_4$ platform does not require phononic engineering of high overtone bulk acoustic resonances or apodization to achieve a flat response. With increased optical mode confinement in lithium niobate, it could be possible to significantly (i.e., more than ten times) increase the LNOD laser tuning efficiency compared to that of piezoelectric AlN actuation \cite{lihachev2021ultralow}.
	
\begin{figure*}[t]
	%	\centering
	%	\includegraphics[width=0.5\textwidth]{figure_1_layout}
	\includegraphics[width=\textwidth]{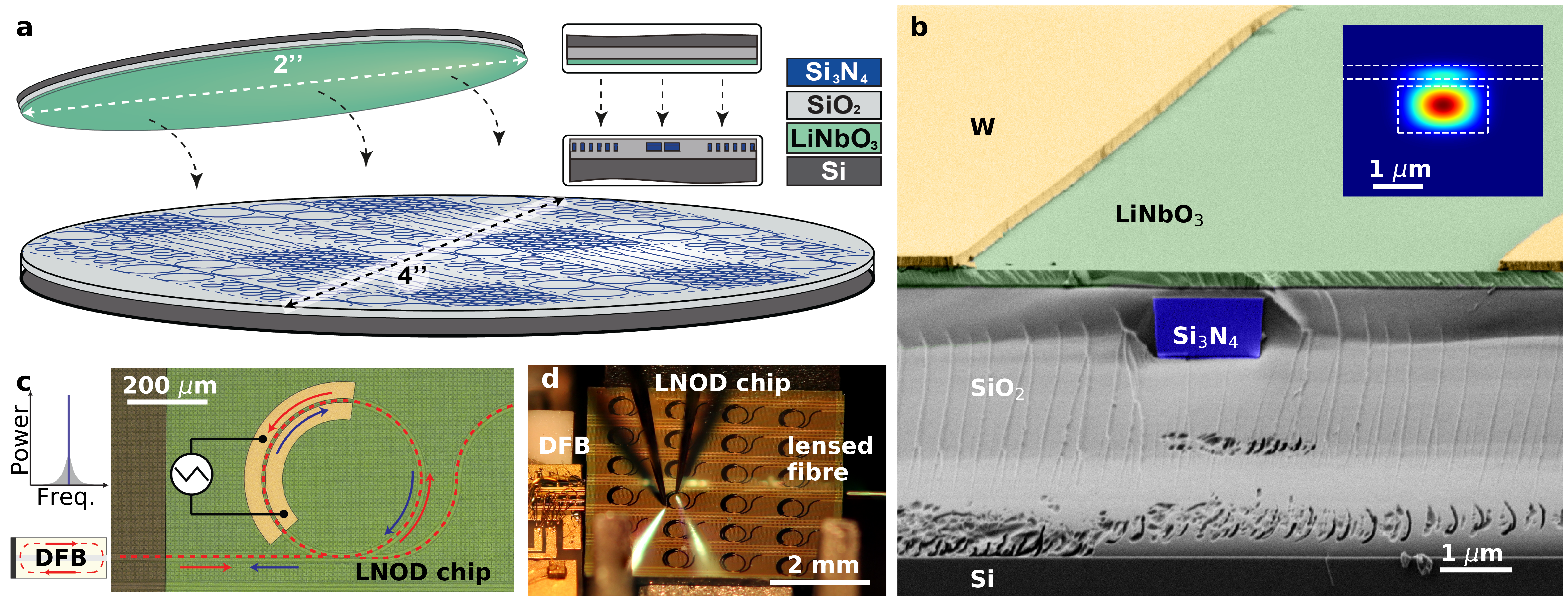}
	\caption{\textbf{Heterogenous $\boldsymbol{\mathrm{Si_3N_4}}$-$\boldsymbol{\mathrm{LiNbO_3}}$ low-loss photonic integrated platform for fast tunable self-injection-locked lasers.} 
	\textbf{(a)} Schematic illustration of lithium niobate on Damascene (LNOD) platform fabrication by heterogeneous integration of a 2'' thin-film lithium niobate wafer onto a 4'' silicon nitride Damascene wafer, with cross-sections of both wafers.  
	\textbf{(b)} False-colour SEM image of an LNOD waveguide cross-section (the original SEM image data is shown in Supplementary Figure 1). The structures in yellow are the tungsten electrodes (W). The inset shows a finite-difference time-domain (FDTD) simulation of the spatial distribution of the hybrid TE-mode electric-field amplitude that was selected for the device design. 
	\textbf{(c)} Schematic illustration of the self-injection locking principle: 
	laser wavelength tuning is achieved by applying a voltage signal (e.g., a linear ramp) on the tungsten electrodes. 
	\textbf{(d)} Photo of the setup with a DFB laser butt-coupled to an LNOD chip. A pair of probes touch the electrodes for electro-optic modulation, and a lensed fibre collects the output radiation.}
	\label{fig_1}
\end{figure*}

\subsection*{RESULTS}

\noindent \textbf{Heterogeneous integration of lithium niobate on silicon nitride.} 
% Our approach is based on a heterogenously integrated $\mathrm{Si_3N_4}$ - $\mathrm{LiNbO_3}$ platform that unifies the individual advantages of $\mathrm{Si_3N_4}$ and $\mathrm{LiNbO_3}$.
% Recent advancements in fabrication of $\mathrm{Si_3N_4}$ PIC have endowed this passive platform with ultralow loss down to 1~dB/m \cite{Liu:21}, high power handling \cite{Brasch:15}, low Raman and Brillouin gains \cite{Karpov:16, Gyger:20}, radiation hardness \cite{Brasch:14}, and high fabrication yield without crack formation \cite{Pfeiffer:18b, Liu:20}.
% In the meantime, LNOI \cite{Zhang:17, Wang2018} has recently emerged as an ideal CMOS-compatible PIC platform featuring high Pockels nonlinearity and low loss.
\noindent Our fabrication method combines the processes of photonic Damascene waveguide fabrication with wafer-scale bonding \cite{Komljenovic:16, Chang:17} to enable electro-optic modulation on passive, ultralow-loss silicon nitride PICs, as schematically depicted in Fig. \ref{fig_1}(a). 
Our process starts with the fabrication of a patterned and planarized silicon nitride substrate using the photonic Damascene process \cite{Pfeiffer:18b, liu2021high}. 
Deep-ultraviolet (DUV) stepper lithography is used to pattern waveguides and microresonators on a silicon substrate with 4~$\mu$m thick thermal wet silicon dioxide (SiO$_2$). 
The pattern is then dry-etched into the silicon dioxide layer to form the waveguide preform, followed by a high-temperature reflow of the waveguide preform \cite{Pfeiffer:18} to reduce surface roughness.
Stoichiometric silicon nitride is deposited by low-pressure chemical vapour deposition (LPCVD) on the patterned substrate, filling the preform and forming the waveguide cores.
Chemical mechanical polishing (CMP) is used to remove excess silicon nitride and to planarize the wafer top surface.
Subsequently the entire substrate is thermally annealed at 1,200$^\circ$C to drive out the residual hydrogen contained in the silicon nitride.
A silicon dioxide interlayer is deposited on the substrate, followed by densification. 
The interlayer is then polished to reduce the remaining topography and set the desired thickness.
A root-mean-square roughness of less than 0.4~nm over an area of a few square micrometres and of only a few nanometres over an area of several hundred micrometres are necessary for bonding.
% Low surface roughness and long-range uniformity are particularly critical for bonding with LNOI; that is why this second CMP step is essential.
Next, a few-nanometre-thick alumina layer is deposited by atomic layer deposition (ALD) on both the donor (LNOI) and the acceptor (Damascene) wafers before contact bonding and donor-wafer removal. 
Tungsten (W) electrodes are then manufactured by sputtering and reactive ion etching. 
At this point, the areas of the coupling facets and the tapered sections of the silicon nitride waveguides are cleared from lithium niobate by physical etching with argon ions, so that light can first couple into the tapered waveguides with minimal coupling losses, before transitioning to the area with lithium niobate bonded on top.
Finally, chip release is performed by chip-facet definition by deep silicon dioxide and silicon etching, followed by chip separation through backside silicon lapping.
Fig.~\ref{fig_1}(b) depicts a scanning electron microscope (SEM) cross section of the heterogeneous LNOD waveguide.
	
\begin{figure*}[t]
		\centering
		\includegraphics[width=0.7\textwidth]{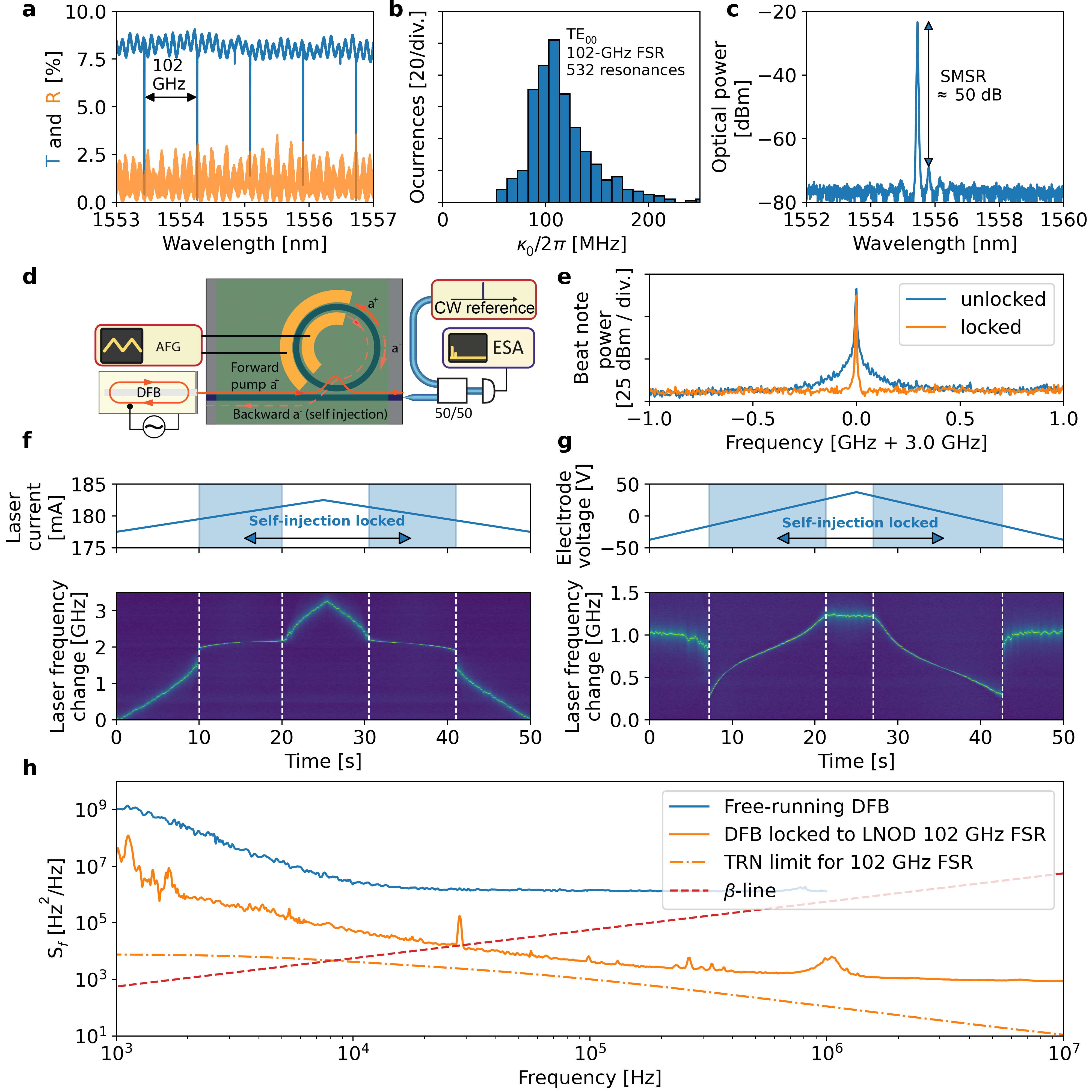}
		\caption{\textbf{DFB laser performance characterization.} 
			\textbf{(a)} Transmission (blue) and reflection (orange) spectra of an LNOD 102-GHz FSR resonator (see Supplementary Figure 2 for full data set). 
			\textbf{(b)} The histogram shows the distribution of the resonance linewidth of 593 resonances of the 102-GHz FSR device with a median linewidth of $\approx100$ MHz, corresponding to a quality factor of $1.9\times 10^6$ ($\kappa_\text{0}$ is the intrinsic cavity decay rate). 
			\textbf{(c)} Optical spectrum of the free-running DFB laser diode. 
			\textbf{(d)} Experimental setup for linewidth measurements with the hybrid integrated laser using the heterodyne beat-note method. 
			\textbf{(e)} Comparison of the laser linewidth for the free-running DFB case and the case where the DFB is self-injection-locked to an LNOD microresonator. 
			\textbf{(f)} Time--frequency map of the beat note showing the laser frequency change on linear modulation of the diode current. The white dashed lines mark the boundaries of the self-injection locking bandwidth, where almost no laser-frequency change is observed.  
			\textbf{(g)} Time--frequency map of the beat note showing the laser-frequency change on linear tuning of the cavity resonance by applying voltage to the electrodes. The DFB current remained fixed in the self-injection locking range.
			\textbf{(h)} Frequency noise spectra of the free-running DFB (blue) and the DFB self-injection-locked to the 102-GHz FSR LNOD resonator (orange). The evaluated thermo-refractive noise limit and the $\beta$-line are given for reference (orange dash-dotted and red dashed lines, respectively).}
		\label{fig_2}
\end{figure*}

\noindent \textbf{Laser self-injection locking.} 
Laser self-injection locking is initiated by butt-coupling of an InP distributed-feedback (DFB) diode laser to the LNOD chip (Fig. \ref{fig_1} (c,d)) and tuning the laser current to match the output frequency to the resonance frequency of the LNOD resonator.
The optical backreflection on surface or volumetric inhomogeneities inside the microresonator provides spectrally narrowband feedback to the laser diode by coupling clock- and counterclockwise propagating modes.
Light in the counterclockwise mode radiates back to the laser bearing the power fraction given by the reflection coefficient $R$, which depends on the interaction strength of the modes and the resonator coupling efficiency \cite{gorodetsky2000rayleigh}. 

%It consists of optical injection due to the narrowband optical feedback induced by a resonant backscattering on the surface and volumetric inhomogeneities of a high-Q resonator (a master oscillator) to which a laser (a slave oscillator) is locked (Fig. \ref{fig_1} (c) and (d)).
% Let us consider the case when a laser with frequency $\omega$ is self-injection locked to one of the microresonator modes with total linewidth $\kappa=\kappa_\text{ex}+\kappa_0$, where $\kappa_\text{ex}$ and $\kappa_0$ are external and intrinsic cavity decay rates, respectively. 
% 
% Rayleigh scattering leads to the generation of an optical mode counter-propagating with respect to the one that couples into the resonator.

The laser diode is forced to oscillate at the frequency of the cavity resonance in the self-injection-locked regime. Assuming that the frequency noise of the laser is white, the frequency noise suppression ratio is \cite{kondratiev2017self}:

%The coefficient of reflection $R$ enabled by outcoupling of the mode counter-propagating with respect to the one that couples into the resonator, and generated because of the Rayleigh scattering with intermodal coupling rate of $\gamma$, is given by the following relation \cite{raja2019electrically}:
%
%\begin{equation}
%	R \approx \frac{2 \eta \Gamma}{1+\Gamma^2},
%\end{equation}

%\noindent where $\eta=\frac{\kappa_{ex}}{\kappa}$ and $\Gamma=\frac{\gamma}{\kappa}$ are dimensionless parameters that represent the resonator coupling efficiency and the modes interaction strength. 

\begin{equation}\label{narrowing}
	\frac{\delta \omega}{\delta \omega_\text{free}} \approx  \frac{Q_\text{DFB}^2}{Q^2}\frac{1}{16 R(1+\alpha_g^2)},
\end{equation}

\noindent where $\delta \omega_\text{free}$ is the linewidth of the free-running DFB laser; $\delta \omega$ is the linewidth of self-injection-locked DFB laser; $Q_\text{DFB}$ and $Q=\omega/\kappa$ are the quality factors of the laser diode cavity and of the microresonator mode, respectively (with $\kappa=\kappa_\text{ex}+\kappa_0$, where $\kappa_\text{0}$ and $\kappa_\mathrm{ex}$ are the intrinsic cavity decay rate and bus--waveguide coupling rate); and $\alpha_g$ is the phase-amplitude coupling factor \cite{henry1982theory}. Self-injection locking occurs within a finite frequency interval around the cavity resonance. 
The locking bandwidth $\Delta \omega_\text{lock}$ is given, assuming large intermodal interaction strength and high coupling efficiency, by:

\begin{equation}\label{range}
	\Delta \omega_\text{lock} \approx \sqrt{R(1+\alpha_g^2)} \frac{\omega}{Q_\text{DFB}}
\end{equation}

In order to strongly reduce the laser linewidth and increase the frequency locking range, a high-Q resonance and strong reflection are desirable.
The device used in our experiments features a 102-GHz free spectral range (FSR) and a resonance linewidth of 100~MHz (see Fig. \ref{fig_2}(a,b)) operated close to critical coupling.
The intrinsic loss of the microresonator indicates a linear propagation loss of our LNOD waveguide platform of 8.5~dB/m.
The power reflection of the device reaches 3\% (see Fig. \ref{fig_2}(a) and Supplementary Figure 2 for full spectrum) and features both the narrowband reflection of the microresonator and wideband sinusoidal modulation by spurious reflections from the chip facet, as well as and the transitions between the inverse tapers the LNOD waveguide.
Despite the weak back-reflection contrast, injection locking is observed due to the narrow linewidth of the optical resonance.
The self-injection-locked DFB emission spectrum (Fig.~\ref{fig_2}(c)) indicates a lasing wavelength of 1555.4~nm with a side-mode-suppression ratio of 50~dB. 
To characterize self-injection locking, a heterodyne beat note of the unlocked and locked DFB laser with a reference laser (Toptica CTL) is generated on a fast photodiode and processed using an electrical spectrum analyzer (Fig.~\ref{fig_2}(d)). 
We observe the narrowing of the beat note upon locking of the DFB laser (see~Fig.~\ref{fig_2}(e)). 
When sweeping the DFB current, we found the regions where there is almost no change of the laser frequency, because of the self-injection locking (see Fig.~\ref{fig_2}(f)). 
To reveal the locking bandwidth, we set the DFB current inside the locked state, and scanned the cavity resonance by applying a triangular voltage chirp to the electrodes (Fig.~\ref{fig_2}(g)). 
The self-injection locking is achieved within a frequency span of around 1~GHz; however, linear tuning is observed only within a 600-MHz band, due to the low back-reflection of the LNOD microresonator. 
Next, we measured the frequency noise single-sideband spectral density $S_{ff}(f)$ of the DFB diode laser in the free-running and self-injection-locked regimes (see Methods for details and Figure \ref{fig_2}(h) for results). 
The laser self-injection locking suppresses the frequency noise by at least 20~dB across all frequency offsets. 
%To calculate the laser linewidth we integrate the frequency noise spectra from 10 Hz offset up to the offset corresponding to the point of the intersection with the beta-line \cite{di2010simple} $S_f(f)=8\ln2\cdot f/\pi^2$. The area under the curve $A$ is then recalculated to provide the full-width-half-maximum (FWHM) measure of the linewidth using the following expression: FWHM $=\sqrt{8 \ln2 \cdot A}$. 
We find the intersection point of the frequency noise curve and the beta-line \cite{di2010simple} at 30~kHz (Fig.~\ref{fig_2}(h) orange and dashed red). 
The full width at half maximum linewidth, which is calculated by integration of the frequency noise from the beta-line to the inverse integration time, is 56~kHz at 0.1~ms integration time, 262~kHz at 1~ms, and 1.1~MHz at 100~ms. 
The laser frequency noise reaches a horizontal plateau (white noise floor) of 10$^3$~Hz$^2$/Hz at 3~MHz offset, which corresponds to an intrinsic laser linewidth of 3.14~kHz. 
%before rising at offsets around 1~MHz due to photon shot noise ($S_\mathrm{ff}(f) \propto f^2$). 
\\

\begin{figure*}[t]
	\includegraphics[width=\textwidth]{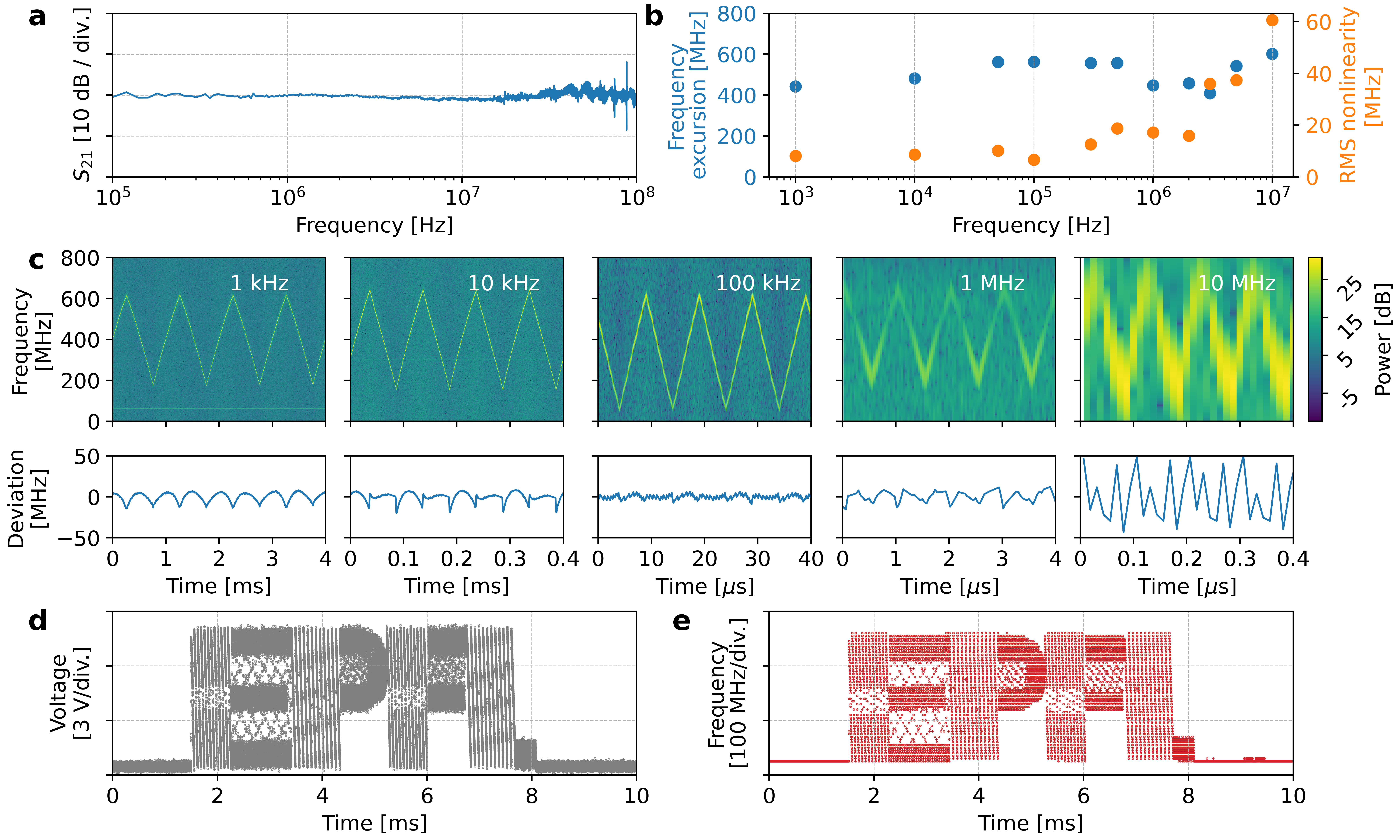}
	\caption{\textbf{Self-injection-locked DFB electro-optic frequency tuning.} 
		\textbf{(a)} Measured response of the electro-optic modulation for the LNOD device utilizing tungsten electrodes. 
		\textbf{(b)} Frequency excursion (blue) and the absolute deviation of the measured tuning profile from a perfect triangular ramp (orange). The deviation was calculated as the difference between the experimental data and the least-squares fitting. 
		\textbf{(c)} Upper row: time--frequency spectrograms of heterodyne beat note for modulation frequencies 1 kHz -- 10 MHz. Lower row: the deviation of the experimental tuning data from the least-squares fit for the same modulation frequencies. 
		\textbf{(d)} Voltage profile applied to the electrodes from an arbitrary waveform generator, resembling the EPFL logo. 
		\textbf{(e)} Measured laser heterodyne beat note showing laser-frequency evolution in the form of the EPFL logo at 450~THz/s tuning rate.}
	\label{fig_3}
\end{figure*}

% The demonstration of the ability to tune the laser diode frequency in arbitrarily complicated way at a high tuning rate. Upper row: the time-frequency plot showing the laser frequency rapid tuning in the pattern resembling the EPFL university logo at the maximum tuning rate of 450 THz/s with the repetition frequency of 76 Hz. Lower row: the corresponding voltage profile generated by an arbitrary waveform generator and applied to the tungsten electrodes on a LNOD chip resonator device. \textbf{(e)} The same as in \textbf{(d)}, but for the IBM corporation logo at the higher tuning rate 550 THz/s with the higher repetition frequency of 760 Hz.
	
\noindent \textbf{Frequency-agile laser tuning.} In order to measure the voltage-to-frequency response of the LNOD microresonator, the signal from a network analyser was applied to the electrodes and the laser frequency (Toptica CTL) was fixed on the slope of the cavity resonance. 
This measurement reveals an important advantage of the LNOD platform: the modulation response function for the 102~GHz FSR microresonator is flat up to 100~MHz (see Fig. \ref{fig_3}(a)). 
To demonstrate the frequency agility of the laser and the response of the laser frequency to a large-amplitude voltage modulation, the DFB laser was self-injection-locked to the cavity resonance and a triangular voltage signal with and modulation frequencies ranging from 1~kHz to 10~MHz was applied. 
No signal pre-distortion \cite{zhang2019laser} or active feedback \cite{roos2009ultrabroadband} were applied to the driving signal.
The applied voltage modulates the refractive index of lithium niobate via the Pockels effect and shifts the cavity resonance, forcing the laser to follow the resonance as long as it remains within the overall locking range \cite{ilchenko2011compact,lihachev2021ultralow}.
To reveal the time-varying frequency-tuning characteristics for large signal modulation in the self-injection-locked state, the heterodyne beat note of the hybrid integrated laser with the reference laser was recorded on a fast photodiode. 
Frequency excursions remained on the level of 500~MHz, independently of the modulation frequency, whereas the nonlinearity tended to increase with increased the modulation frequency. 
The minimum nonlinearity of 1\% of the frequency excursion is observed at a 100~kHz tuning rate. 
The upper row in Fig.~\ref{fig_3}(c) presents the processed laser-frequency spectrograms, which are calculated by time-segmented Fourier transformation, and the bottom row shows corresponding residuals after a perfect triangular modulation is fit to the data.
Figure~\ref{fig_3}(b) shows the laser frequency excursion and the root-mean-squared deviation of the measured profiles from a perfect triangular frequency modulation determined by curve fitting. Additional data on tuning efficiency and hysteresis are shown in Supplementary Figures 3 and 4. 
The demonstrated frequency-modulation rate of 600~MHz in 50~ns equates to 12~PHz/s.

Although highly linear ramp frequency modulation is essential for FMCW LiDAR application \cite{uttam1985precision}, the frequency can be modulated in an arbitrary manner while preserving of the high tuning rate. 
To illustrate this, we programmed an arbitrary waveform generator to reproduce the logo of EPFL (see Fig. \ref{fig_3}(d)) and applied the signal to the LNOD device. 
The laser frequency was again determined by heterodyne beat note with the reference laser, and the result of the time--frequency analysis is depicted in Fig. \ref{fig_3}(e).
	
\begin{figure*}[t]
	\centering
	\includegraphics[width=\textwidth]{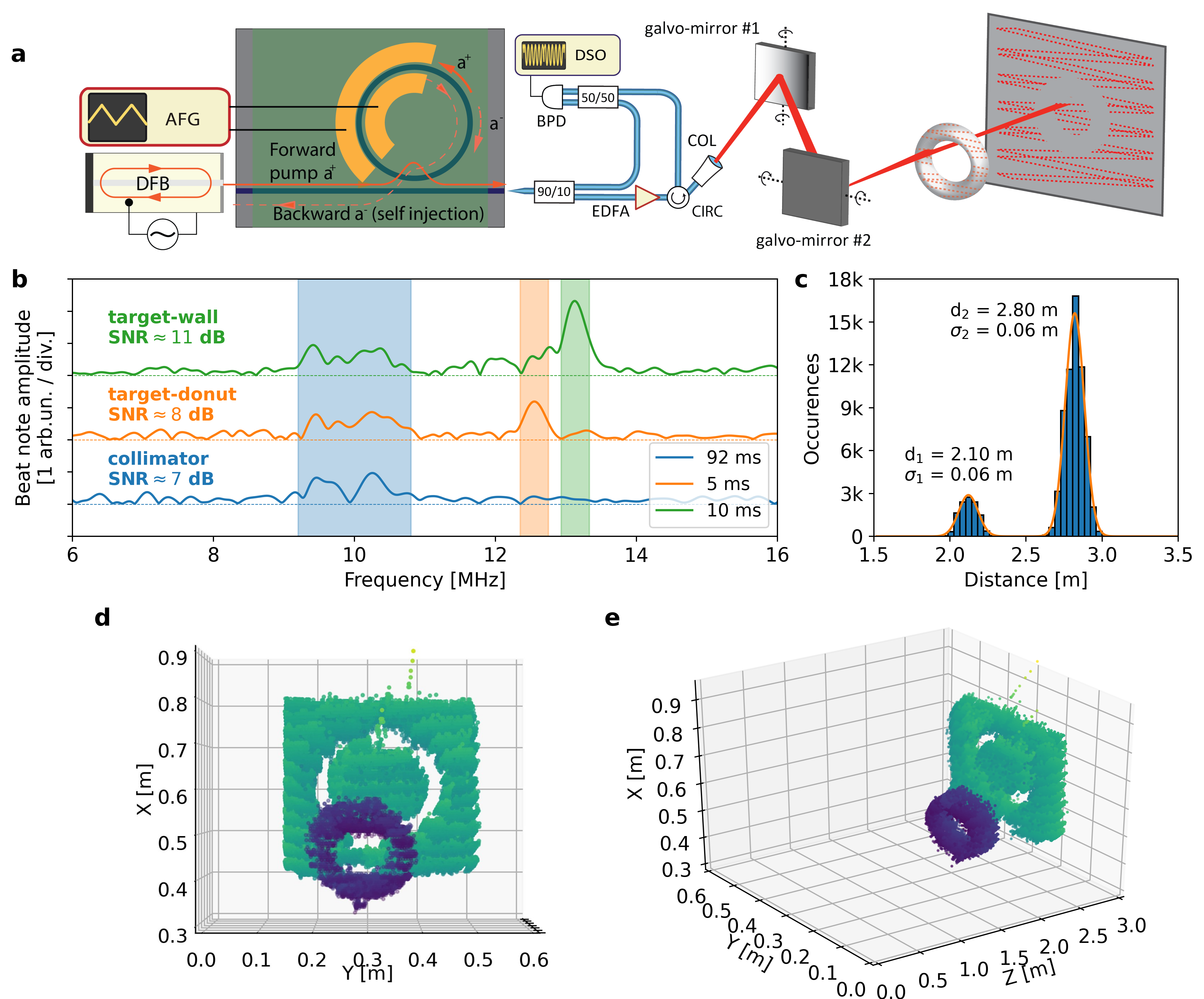}
	\caption{\textbf{Coherent electro-optic LiDAR demonstration.} 
		\textbf{(a)} Schematics of the experimental setup for coherent optical ranging. The output signal of the tunable laser source with a linear frequency chirp is split into two channels for the delayed homodyne detection. The signal in the first channel is amplified and, by means of mechanical beam-steering, scans the target. The signal in the second channel is mixed with the fraction of the power of the first channel that was scattered by the target. The beat note power evolution is recorded by an oscilloscope. AFG, arbitrary function generator; DSO, digital storage oscilloscope; EDFA, erbium-doped fibre amplifier; CIRC, optical circulator; BPD, balanced photodiode; COL, collimator. 
		\textbf{(b)} Examples of the delayed homodyne beat note corresponding to signals from the collimator (blue line), the donut (orange line), the wall (green line) with the respective values of the signal-to-noise ratio. \textbf{(c)} Histogram showing the distribution of the calculated values of distance to the target. The two peaks correspond to the reflections from the donut and the wall. Both peaks are fitted with a double-Gaussian function with fitting parameters, the mean distance and standard deviation, indicated. 
		\textbf{(d,e)} Point-cloud representation of the measured target scene.}
	\label{fig_4}
\end{figure*}
	
\noindent \textbf{Optical coherent ranging demonstration.} In order to demonstrate the application potential of our laser, we perform a proof-of-concept optical ranging experiment inside a lab environment.
The FMCW LiDAR method \cite{Behroozpour2017} consists of a linear frequency modulation of the laser source and delayed homodyne detection with the optical signal reflected from the target. 
The experimental setup is depicted in Fig.\ref{fig_4}(a) (see the Methods section for a detailed description). 
%A linear voltage ramp with 25~V peak-to-peak (Vpp) amplitude and a modulation frequency of 100 kHz is applied to the electrodes. 
%This value of the frequency excursion is equivalent to 20-cm resolution in distance measurements. 
The laser beam is scanned across the target scene by means of two galvo-mirrors with triangular driving signals. 
We used a polystyrene donut-like object and a metal sidewall of a rack enclosure as the target. 
Both objects were located approximately 3~metres away from the collimator. 
A photograph of the target scene and the beam-scanning pattern are depicted in Supplementary Figure 5. 
The beat note between the signal reflected from the target and the local oscillator is detected with a balanced photodiode and recorded by an oscilloscope. 
Zero-padded short-time Fourier transformation is then applied to the collected oscillogram data to retrieve the beat note spectrum evolution over 128k time slices.
The time--frequency spectrograms obtained for both target and the reference Mach--Zehnder interferometer (MZI) are shown in the Supplementary Figure~6. 
The MZI was used for distance calibration (we retrieve the resolution of 15 cm) only and no signal pre-distortion or active feedback was applied. 
Figure ~\ref{fig_4}(b) shows three different timeframes with beat notes of the local oscillator with the reflections from the wall, the donut and the collimator, and their respective signal-to-noise values.
Last, the centre frequencies of the beat note spectra were identified and mapped into the distance domain using the MZI length as a reference.  
%Characteristic peaks in the vicinity of 10 MHz with SNR of 7 dB observed on all the traces showed in Fig. \ref{fig_4}(b) are associated with the reflection from the collimator lens and contain no information about the target elements. 
%Another characteristic peaks observed at higher frequencies - from 12 to 13 MHz - is the actual reflection from the donut and the wall (8 dB and 11 dB, respectively). Searching for the frequency values corresponding to the maximum of the spectra taken at every instant included on the time axis, one finds the points to be mapped from the frequency into the distance domain. The points with SNR values less than 1 dB are included neither for the subsequent steps of data processing and nor for the point cloud representation. 
The resulting distribution of the distance values is plotted as a histogram in Fig. \ref{fig_4}(c), showing two peaks representing the donut (at 2.1~m) and the wall (at 2.8~m).
The double-Gaussian fit reveals the statistical distribution of distance values for both objects (see Fig. \ref{fig_4}(c)).
The point cloud of the 3D optical ranging is inferred from the distance data and the voltage-to-angle conversion of the galvo-mirror controller; it is shown in Fig. \ref{fig_4}(d) and (e), where the point colour encodes the distance from the collimator.

\subsection*{CONCLUSION}
In summary, we have demonstrated a novel heterogeneous platform for electro-optical photonic circuits using heterogeneous integration of photonic Damascene silicon nitride and thin-film lithium niobate with good fibre-to-chip coupling (3.9~dB per facet) and linear propagation losses as low as 8.5~dB/m. 
We have further demonstrated optical microresonators with 100~MHz loaded linewidth, wideband uniform bus--waveguide coupling and flat modulation response up to 100~MHz. 
%Of the many applications that potentially benefit from such a rapidly and precisely tunable photonic integrated platform, we choose to perform a proof-of-principle demonstration of laser self-injection locking with linear tuning at rates as high as 12~PHz/s and simultaneously narrow linewidth \cite{Harris1998}. 
Using the combination of ultra-low-loss silicon nitride photonic circuits, and endowing them with lithium niobate electro-optical modulation enables to create hybrid integrated injection locked lasers with simultaneously narrow linewidth and fast on chip tuning of 12 PHz/second. 
This laser source enables frequency-modulated continuous-wave optical ranging without the need for signal pre-distortion or active feedback, with a resolution of around 15~cm.
With improvements in PIC design, such as a reduction of the interlayer silicon dioxide thickness and electrode-position optimization, we believe that our platform will form the basis of very fast tunable lasers with 10-ns level switching time, mode-hop-free tuning over tens of GHz, and fundamental linewidths below 100~Hz and km-level coherence length. 
Beyond integrated lasers, the platform can also be used to realise other devices, such as photonic switching networks for photonic computing \cite{Drake1986} or boson sampling \cite{B.2013}, as well as integrated transceivers that combine hybrid integrated narrow linewidth lasers with lithium-niobate-based integrated IQ-modulators. The wide transparency window of both lithium niobate \cite{Jhans1986} and silicon nitride \cite{Morin:21,Guo2018}, moreover, allows such frequency agility to be extended to other wavelength ranges such as the mid-infrared or the visible, providing a platform for fast tunable lasers for applications ranging from optical coherence tomography \cite{ji2019chip} to trace gas sensing \cite{Truong2013}.
%Such a laser would bear many advantages for a variety of applications in high-resolution optical ranging and velocimetry \cite{Feneyrou:17,martin2018photonic,rogers2021universal}, rapid trace-gas spectroscopy \cite{debecker2005high,Truong2013,millot2016frequency} and optical communication \cite{Zhang2009}. 
%The wide optical transparency of the involved materials \cite{Jhans1986} will facilitate operation at visible wavelength as well as the near to mid-infrared. 
	
	\begin{footnotesize}
		
		\subsection*{METHODS}
		
\noindent \textbf{Device fabrication.} First, we produce a photonic Damascene silicon nitride integrated circuit. We start with a 4'' silicon (Si) wafer with 4 $\mu$m thermal oxide (SiO$_2$) that we then process as follows \cite{pfeiffer2018photonic}: deep-ultra-violet (DUV) stepper photolithography for preform patterning, preform dry etching, preform reflow process, low-pressure chemical vapour deposition (LPCVD) of silicon nitride (Si$_3$N$_4$), and chemical-mechanical polishing (CMP). 
An SiO$_2$-interlayer is then deposited and subsequently polished with CMP.
Prior to bonding, a few-nanometre-thick alumina layer is deposited on the donor (LNOI) and the acceptor wafer (Damascene).
After that, both wafers are brought in contact and annealed for several hours at 250$^\circ$C. The silicon on the backside of the donor wafer is grinded and what is left after this step is removed with tetramethylammonium
hydroxide (TMAH) wet etching. The thermal oxide is wet etched with buffered hydrofluoric acid (BHF). A layer of tungsten (W) is sputtered on the lithium niobate surface, and the electrode pattern is transferred in this layer via fluoride-based reactive ion etching (RIE). Finally, lithium niobate is etched to open the chip facets areas in order to improve the input coupling of the device on the chips by means of argon ion beam etching. The subsequent chip release is performed in three steps: dry etching of chips boundaries in silicon dioxide with fluorine-based chemistry, further etching of silicon carrier by the Bosch process, and backside wafer grinding.    

\noindent \textbf{Laser frequency noise measurements.} We performed heterodyne beat note spectroscopy \cite{duthel2009laser} beating the reference external cavity diode laser (Toptica CTL) with the hybrid integrated DFB diode laser to reveal the frequency noise of the latter. The beat note of the two signals was detected on a photodiode, and its electrical output was then sent to an electrical spectrum analyser (Rohde \& Schwarz FSW43). The recorded data for the in-phase and quadrature components of the beat note were processed by Welch's method \cite{welch1967use} to retrieve the single sided phase noise power spectral density (PSD) $S_{\phi \phi}$ that was converted to frequency noise $S_{ff}$ using: $S_{ff} = f^2 \cdot S_{\phi \phi}$. To calculate the laser linewidth, we integrate the frequency noise spectra from the intersection of the PSD with the beta-line $S_f(f)=8\ln2\cdot f/\pi^2$ down to the integration time of measurement \cite{di2010simple}. The area under the curve $A$ is then recalculated to provide the full-width-half-maximum (FWHM) measure of the linewidth using: FWHM $=\sqrt{8 \ln2 \cdot A}$. Because a rigorous definition of the optical linewidth does depend on the integration time of the measurement we evaluate the FWHM linewidth as 56~kHz at 0.1~ms integration time, 262~kHz at 1~ms, and 1.1~MHz at 100~ms. 
%Extended data of the PSD from offset frequencies starting at 10~Hz is shown in the SI. 
The phase noise of the reference laser is determined by another beatnote measurement with a commercial ultrastable laser (Menlo ORS).
		
\noindent \textbf{Thermo-refractive noise simulations.} The frequency stability of the 102-GHz-FSR LNOD devices is primarily limited by the material refractive index fluctuations due to the relatively large material temperature fluctuations at resonator scale (thermo-refractive noise). To quantify the noise level in our system, we follow the approach based on the Fluctuation-Dissipation theorem (FDT), described in Ref. \cite{levin1998internal,kondratiev2018thermorefractive,huang2019thermorefractive}, that was originally given by Levin and successfully applied to the thermal noise analysis of LIGO's mirrors. Since FDT relates fluctuations of a system to how the system dissipates energy, we simulate the noise levels with finite element method by testing how the system dissipates in response to a probe force. Since the fractional thermo-refractive noise $\frac{\delta \omega}{\omega} = \int d\vec{r} q(\vec{r}) \delta T(\vec{r})$ of our device is a weighted average of the temperature fluctuations $\delta T(\vec{r})$ determined by the optical field distribution $\vec{e}(\vec{r})$, in order to find out its magnitude at a particular Fourier frequency $f$, we apply a sinusoidal entropy oscillation (energy-conjugated with temperature) at this frequency, with the same weight $q(\vec{r})$ mimicking the field distribution, to our system in the simulation. The corresponding power dissipated $W_\text{diss}$ in the system is retrieved from the simulation and is used to calculate the thermo-refractive noise power spectral density $S_{\frac{\delta \omega}{\omega}}(f)$ at this particular frequency using FDT. The device field distribution and the heat propagation simulated in the described steps are performed on COMSOL Multiphysics. 
		
\noindent \textbf{Coherent ranging experiment.} The laser diode is edge-coupled to the LNOD chip with 200-nm-thick tungsten electrodes deposited along the silicon nitride waveguide on lithium niobate. The laser frequency tuning is achieved by locking the laser to a cavity resonance, fixing the DFB current, and tuning the cavity resonance via Pockels effect by a voltage applied to the electrodes. Triangular ramp signal from AWG with 0.5 Vpp amplitude and 100 kHz frequency is further amplified up to 25 Vpp by a high-voltage amplifier (Falco Systems) with 5 MHz bandwidth. No additional pre- or post- processing (linearization) was utilized for the laser frequency ramp for the coherent ranging experiment. 
%We used the cavity resonance corresponding to 179 mA DFB current and measured 0.7 GHz laser frequency excursion. It is equivalent to the distance resolution of 20 cm. 
We used the cavity resonance corresponding to 179 mA DFB current.
To calibrate the frequency excursion, the 5 \% fraction of the optical signal was sent to a reference Mach-Zehnder (MZI) fibre interferometer. The MZI optical length of 13.18 m was found out by an independent measurement involving a tunable diode laser scan calibrated by a frequency comb. Taking the measured MZI optical length and beatnote frequency values, the distance resolution of 15 cm is inferred. 95 \% of light is split into two paths: the local oscillator path (10 \%) and the target path (90 \%). The signal in the target path is amplified by an EDFA (Calmar) up to 4 mW and directed to the collimator with the 8-mm aperture set to match the target distance range of 3~m. We use the galvo scanner (Thorlabs GVS112) for the beam steering. Two mirrors were controlled by linear ramp signals of 3~Hz and 60~Hz rates with the amplitude and offset values chosen to ensure that the scanning pattern fully covers the target scene. The data for the point cloud was collected within the total time interval of 1.3~s. The frame rate was limited by the galvo scanning speed and the Doppler broadening that is imparted by the rapidly tilting mirrors.
		
\noindent \textbf{Scene reconstruction and signal data processing.} The data collected in the FMCW LiDAR experiment were subject to digital signal processing steps in order to locate the scene elements in space. First, the zero-padded short-time Fourier transforms (STFT) of the beat note oscillograms from the target and the reference MZI were evaluated. The Blackman-Harris window function was used with the window size set to one period of the frequency modulated signal. Second, the obtained time-frequency maps were used to search at any given time frame the frequency values corresponding to the beat note peak. This set was filtered so that only the data points with beat note amplitudes above 1.3 were considered for further analysis. We then subtract the distance from the laser to the collimator so that the point cloud distance is given with respect to the collimator aperture position. Finally, the frequency data were converted to the distance domain, using MZI length as a reference, and the Cartesian components of each point were computed from the voltage profile applied to the galvo-mirrors.
				
\noindent \textbf{Electro-optic S$_{21}$ response measurement.} A 300 $\mu$W continuous wave (CW) light (1550 nm) from an external cavity diode laser (Toptica CTL) is  coupled to the device using a lensed fibre. The input laser is biased at the slope of the optical resonance. A -5 dBm RF electrical signal is applied from port 1 of the network analyzer to the electrodes of the device under test, and the light intensity modulation is detected by a 12~GHz photodiode (New Focus 1544), which is sent back to port 2 of the network analyzer.
		\\
%		\\
		
\noindent \textbf{Authors Contributions}:
V.S. and G.L. performed experiments with the help of J.R. and A.S.. 
R.N.W., A.R., C.M., and J.L. developed the processes and fabricated the samples with assistance from S.H..
M.C. designed the lithography masks and performed PIC simulations.
U.D., Y.P., A.R., R.N.W., and J.L. performed the CMP for bonding. 
D.C. performed the wafer bonding. 
V.S., G.L., and J.R. analysed the data. 
V.S. and G.H. performed thermo-refractive noise limit simulations.
V.S., G.L., J.R., and T.J.K. wrote the manuscript with input from A.R., A.S., and P.S. 
P.S. and T.J.K. supervised the project.

\noindent \textbf{Funding Information}:
This work was supported by Contract HR0011-20-2-0046 (NOVEL) from the Defense Advanced Research Projects Agency (DARPA), Microsystems Technology Office (MTO), and by funding from the European Union Horizon 2020 Research and Innovation Program under the Marie Sklodowska-Curie grant agreement No. 722923 (OMT) and No. 812818 (MICROCOMB), and under the FET-Proactive grant agreement No. 732894 (HOT). It was also supported by funding from the Swiss National Science Foundation under grant agreement No. 176563 (BRIDGE) and No. 186364 (QuantEOM).% and by the Air Force Office of Scientific Research (AFOSR) under award number FA9550-19-1-0250.
		
\noindent \textbf{Acknowledgments}:
The samples were fabricated in the EPFL center of MicroNanoTechnology (CMi)  and the Binnig and Rohrer Nanotechnology Center (BRNC) at IBM Research. We thank the Cleanroom Operations Team of the BRNC, especially Diana Davila Pineda and Ronald Grundbacher for their help and support.
		
\noindent \textbf{Data Availability}:
The code and data used to produce the plots within this work will be released on the repository \texttt{Zenodo} upon publication of this preprint.
		
\end{footnotesize}
%\bibliographystyle{apsrev4-2}
%\bibliography{EO_LiDAR_ref}

\clearpage
\onecolumngrid

\begin{center}
	\large{\textbf{Supplementary Information to:\\Ultrafast tunable lasers using lithium niobate integrated photonics}}\\[1em]
	\normalsize
	Viacheslav Snigirev$^{1,*}$, Annina Riedhauser$^{2,*}$, Grigory Lihachev$^{1,*}$, Johann Riemensberger$^{1}$, Rui~Ning~Wang$^{1}$, Charles M\"ohl$^{2}$, Anat Siddharth$^1$, Guanhao Huang$^1$, Youri Popoff$^{2,3}$, Ute Drechsler$^2$, Daniele Caimi$^2$, Simon H\"onl$^2$, Junqiu Liu$^1$, Paul Seidler$^{2}$, and Tobias J. Kippenberg$^{1}$\\[0.5em]
	\footnotesize
	$^1$ \textit{Institute of Physics, Swiss Federal Institute of Technology Lausanne (EPFL), CH-1015 Lausanne, Switzerland}\\
	$^2$ \textit{IBM Research Europe, Zurich, Säumerstrasse 4, CH-8803 Rüschlikon, Switzerland}\\
	$^3$ \textit{Integrated Systems Laboratory, Swiss Federal Institute of Technology Z\"{u}rich (ETH Z\"{u}rich), CH-8092 Z\"{u}rich, Switzerland}\\
	$^*$ These authors contributed equally to this work\\
%	$^\dagger$ pfs@zurich.ibm.ch\\
%	$^\ddagger$ tobias.kippenberg@epfl.ch
\end{center}

	\begin{figure*}[h]
	\centering
	\includegraphics[width=0.6\textwidth]{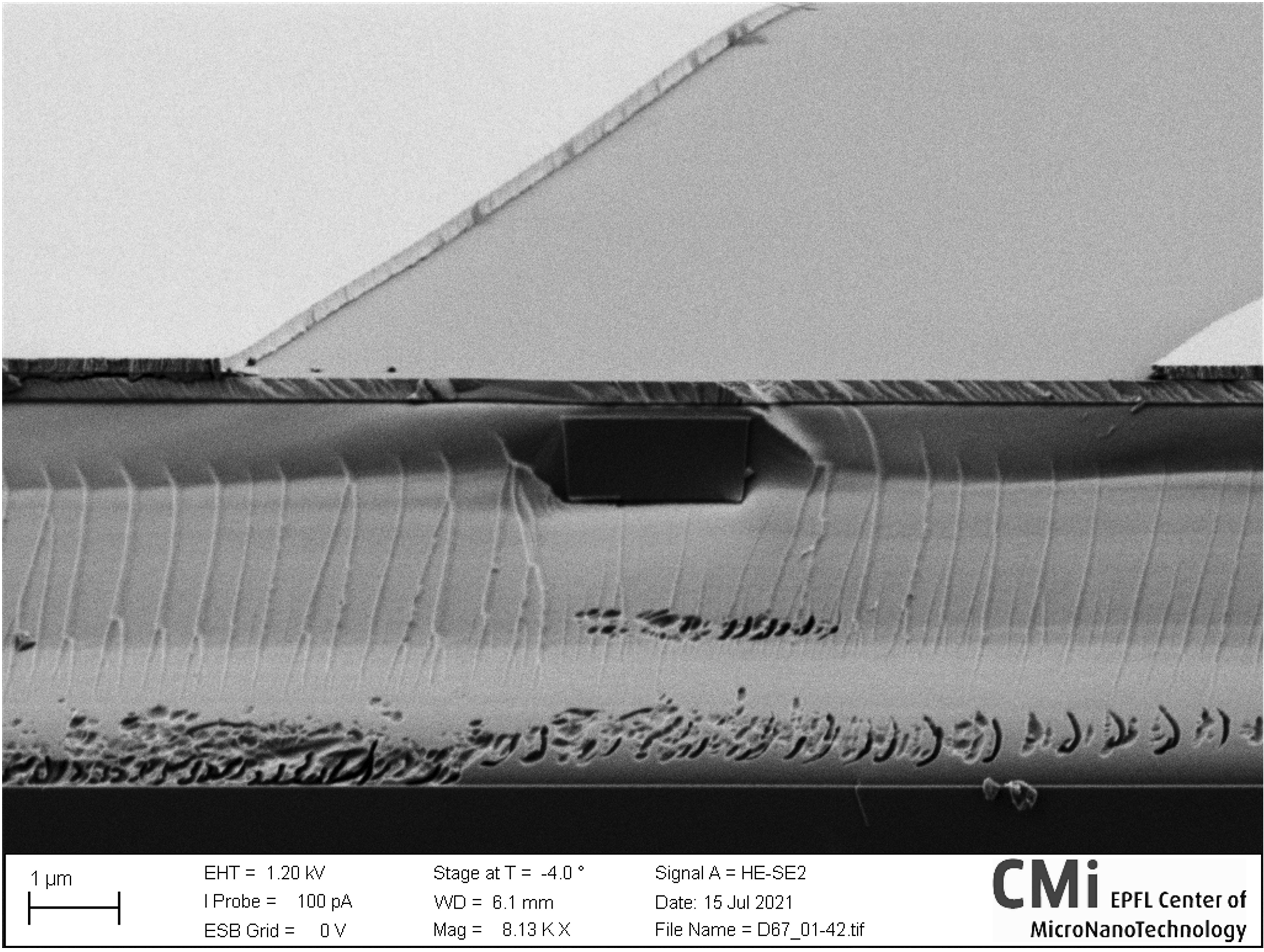}
	\caption{\textbf{Scanning electron microscopy image of a LNOD waveguide.} Original, unprocessed SEM data used to prepare Fig.1b of the main text.}
\end{figure*}

\begin{figure*}[h]
	\centering
	\includegraphics[width=0.7\textwidth]{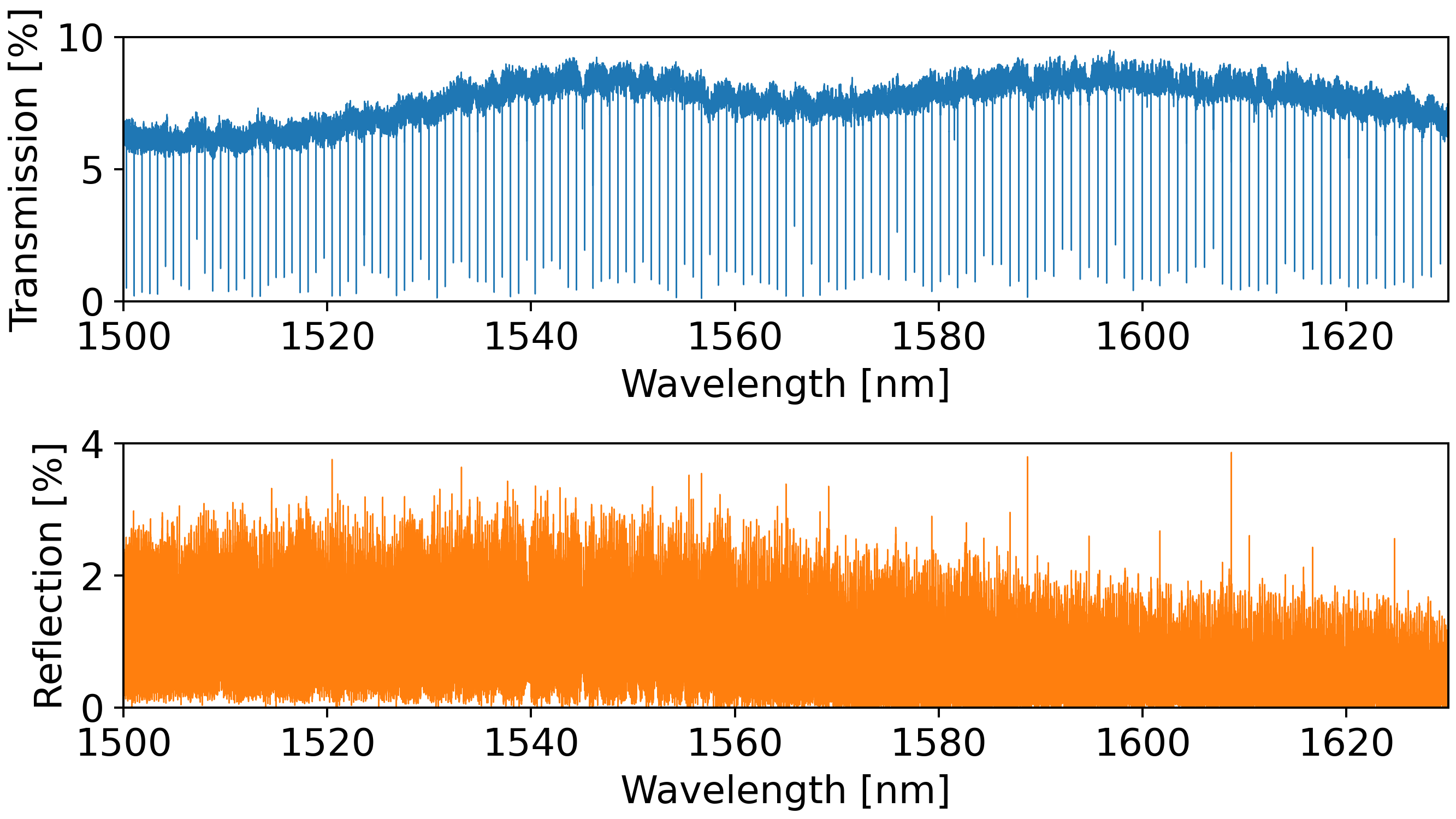}
	\caption{\textbf{Transmission and reflection of a 102-GHz FRS LNOD microring resonator.} \textbf{(Top)} Transmission spectrum: almost all the presented mode resonances are critically coupled. \textbf{(Bottom)} Reflection spectrum. (full measurements span).}
\end{figure*}

%	\begin{figure*}[h]
%		\centering
%		\includegraphics[width=0.7\textwidth]{si_noise_1.png}
%		\caption{Frequency noise spectra of the DFB (full measurements span).}
%	\end{figure*}

%	\newpage

%	\section*{S2: DFB diode laser frequency tuning}

%	\begin{figure*}
%		\centering
%		\includegraphics[width=0.5\textwidth]{setup_heterodyne_with_DSO.png}
%		\caption{The heterodyne spectroscopy setup used for the linear ramp DFB frequency tuning measurements.}
%	\end{figure*}

\begin{figure*}[h]
	\centering
	\includegraphics[width=0.9\textwidth]{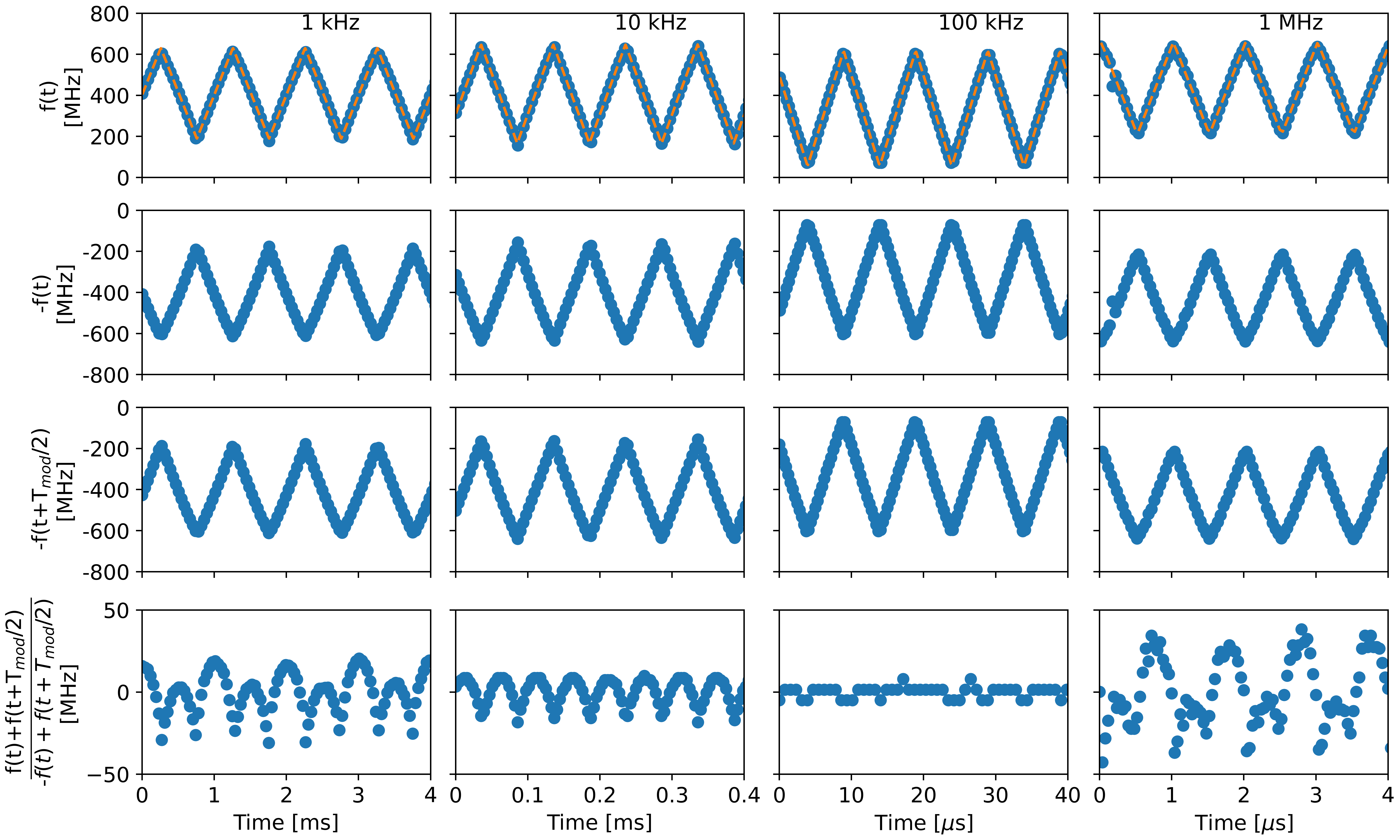}
	\caption{\textbf{Hysteresis evaluation for the DFB laser diode frequency tuning.} The first row represents the heterodyne beatnote evolution for the repetition frequency values of 1 kHz, 10 kHz, 100 kHz and 1MHz, and its fit with a perfect triangular ramp. The second row is the same data, but mirrored with respect to a horizontal axis of 0 MHz. In the third row, there is the next transformation step when the data is moved on a half a period to the left, so that what has been an up-slide in the first row becomes down-slide in the third and vice versa. In the last row, adding the data patterns of the first and the third row and subtracting the mean value from the sum, one observes the hysteresis-induced deviations between the up-slide and down-slide fragment of the ramps.}
\end{figure*}

\begin{figure*}[h]
	\centering
	\includegraphics[width=0.9\textwidth]{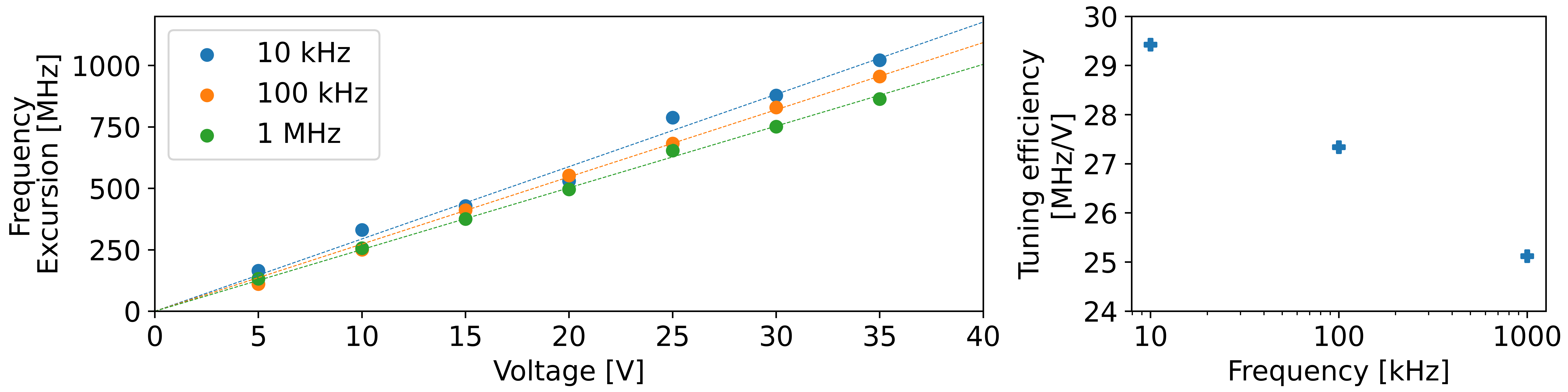}
	\caption{\textbf{Tuning efficiency of LNOD samples.} \textbf{(Left)} Applying a triangular ramp voltage waveform to LNOD device's electrodes with selected values of the modulation frequency (10 kHz, 100 kHz and 1 MHz) and gradually increasing the signal peak-to-peak amplitude, one observes linear growth of the DFB frequency excursion. \textbf{(Right)} To retrieve the tuning efficiency values [MHz/V], the linear model fit of the data can be performed in the voltages diapason where the excursion is smaller than the locking bandwidth limit of $\sim$1 GHz (results are given as the inset plot).}
\end{figure*}

%	\begin{figure*}
%		\centering
%		\includegraphics[width=\textwidth]{freq_tuning_2.png}
%		\caption{Linear triangular ramp frequency tuning at low frequencies.}
%	\end{figure*}
%
%	\begin{figure*}
%		\centering
%		\includegraphics[width=\textwidth]{freq_tuning_1.png}
%		\caption{Arbitrary waveform generator voltage signals programmed to reproduce the logos of EPFL and IBM.}
%	\end{figure*}

%	\newpage

%	\section*{S3: FMCW LiDAR ranging experiments}

\begin{figure*}[h]
	\centering
	\includegraphics[width=0.75\textwidth]{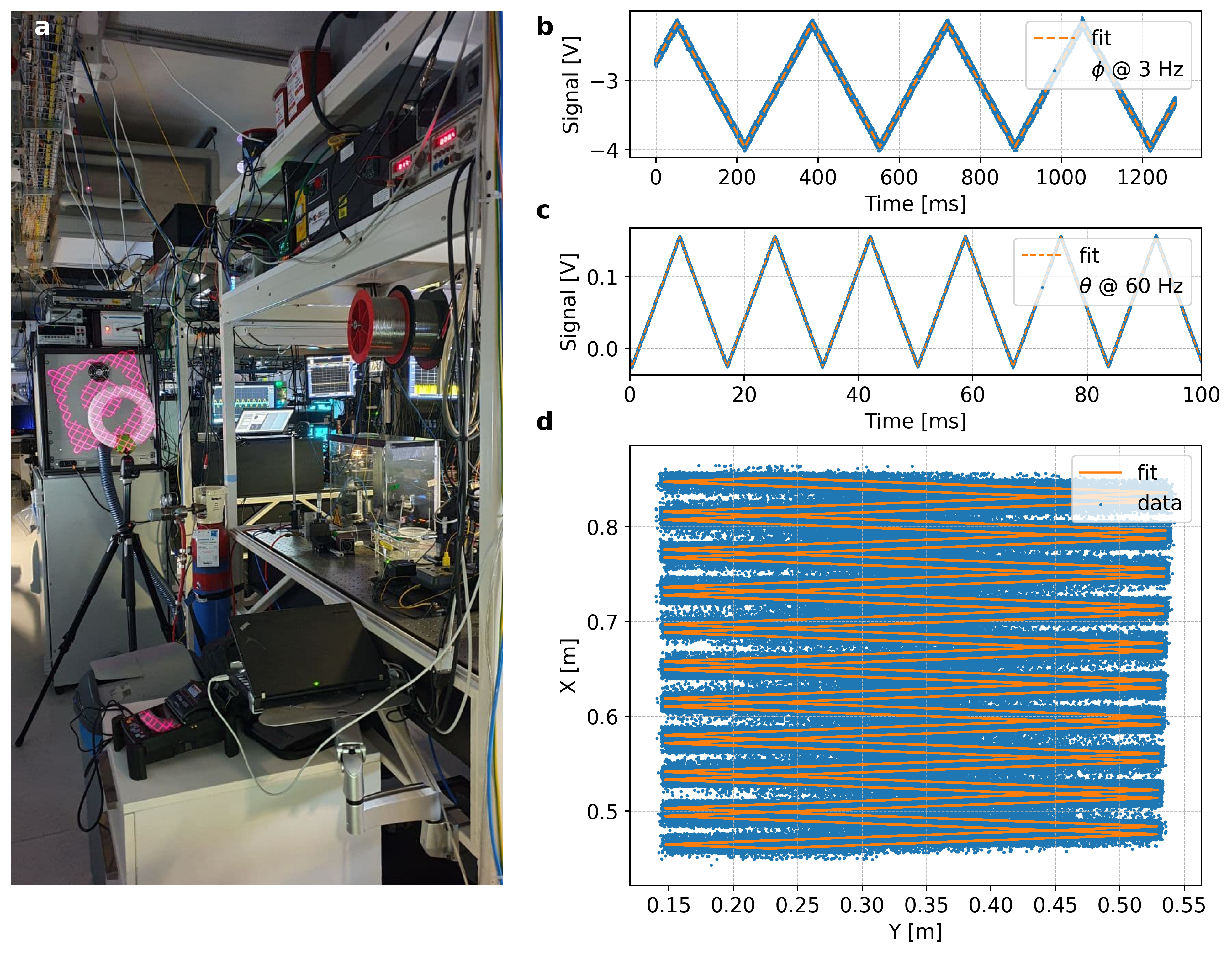}
	\caption{\textbf{FMCW LiDAR setup.} \textbf{(a)} The photo of the setup for coherent optical ranging experiments containing the target - a polystyrene donut mounted on the stage and an instrument box wall behind - and the scanning pattern of the galvo-mirrors embracing all of the target elements. \textbf{(b)},\textbf{(c)} The voltage signal profiles applied to the the two galvo-mirrors enabling two angular degrees of freedom - $\phi$ and $\theta$ - for scanning, and their fits with a perfect triangular ramp. \textbf{(d)} The actual data of scanning pattern and its reconstruction after the fitting the angular coordinates.}
\end{figure*}

\begin{figure*}[h]
	\centering
	\includegraphics[width=0.9\textwidth]{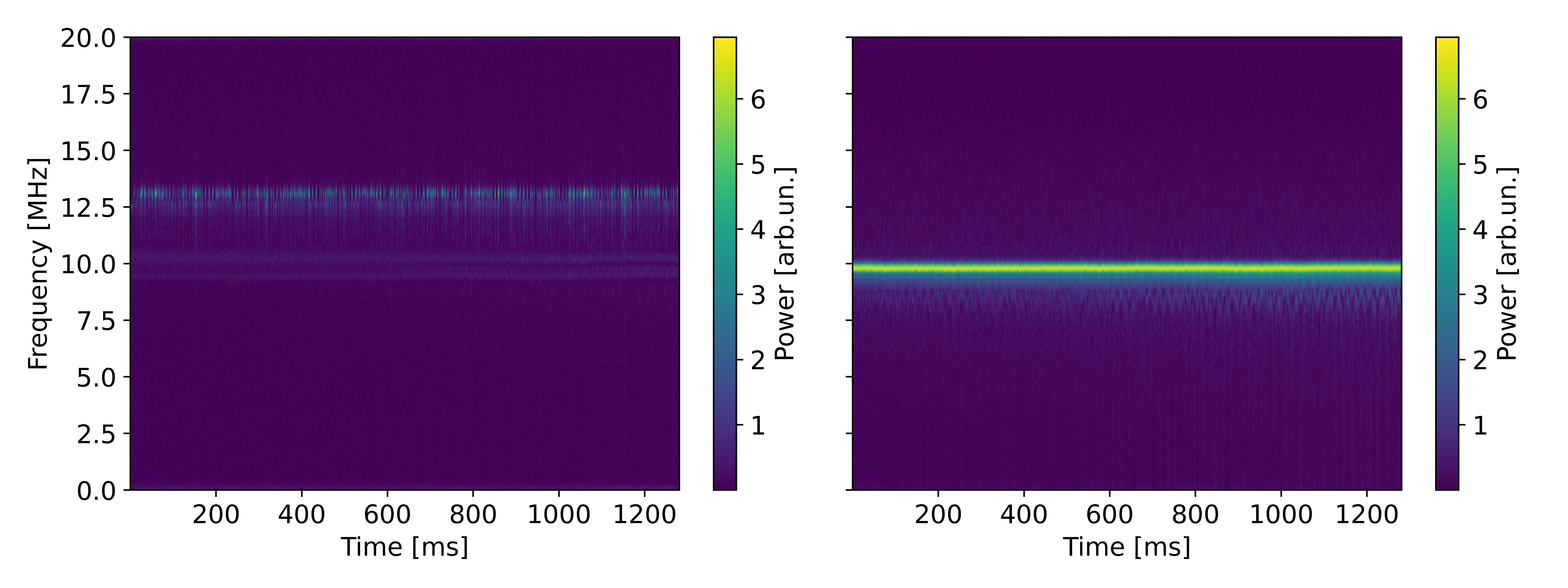}
	\caption{\textbf{Short time Fourier transform of the delayed homodyne beat note oscillogram.} \textbf{(Left)} Time-frequency map calculated for the target response. \textbf{(Right)} The same as in \textbf{(Left)}, but for the reference Mach-Zehnder interferometer. }
\end{figure*}

\end{document}